\documentclass[aps,prb,twocolumn,floatfix,amssymb,superscriptaddress]{revtex4}

\RequirePackage{color}
\usepackage{ulem}

\usepackage{amsmath}
\usepackage{graphicx}

\newcommand{\bk}{\mathbf{k}}

\newcommand{\pdag}{\phantom{\dag}}

\begin{document}

\title{Conductance signatures of Kondo splitting and quantum criticality in triple quantum dots}
\title{Zero-field splitting of the Kondo resonance and quantum criticality in triple quantum dots}
\date{\today}

\author{Arturo Wong}
\author{Francisco Mireles}
\affiliation{Depto. de F\'isica Te\'orica, Centro de
Nanociencias y Nanotecnolog\'ia, Universidad Nacional Aut\'onoma
de M\'exico, Apdo. Postal 2681,
22800 Ensenada, Baja California, M\'exico}

\begin{abstract}
We consider a triple-quantum-dot (TQD) system composed by an interacting quantum dot connected to two effectively non-interacting dots, which in turn are both connected in parallel to metallic leads.  As we show, this system can be mapped onto a single-impurity Anderson model with a non-trivial density of states.   The TQD's transport properties are investigated under a continuous tuning of the non-interacting dots' energy-levels, employing the Numerical Renormalization Group technique.  Interference between single and many-particle resonances splits the Kondo peak, fulfilling a generalized Friedel sum rule. In addition, a particular configuration in which  one of the non-interacting dots is held out of resonance with the leads allows to access a pseudogap regime where a Kosterlitz-Thouless type quantum-phase-transition (QPT) occur, separating the Kondo and non-Kondo behavior.  Within this same configuration, the TQD exhibits traces of the Fano-Kondo effect, which is in turn, strongly affected by the QPT. Signatures of all these phenomena are neatly displayed by the calculated linear conductance. 
\end{abstract}

\pacs{72.15.Qm,73.63.Kv,73.23.-b,64.70.Tg}

\maketitle

\section{Introduction}

	Semiconductor quantum dots (QDs) constitute an exciting playground to test strongly-correlated phenomena, due to their high tunability through external voltages.  Specifically they offer a unique environment to study one of the most fundamental many-body interactions: the Kondo effect.\cite{Hewson, Kouwenhoven} This phenomenon has been widely investigated in QD systems, both experimentally\cite{Goldhaber-Gordon, Cronenwett, Jeong:2001, Chen:2004, Craig:2004, Hubel:2008} and theoretically.\cite{Glazman:1988, Costi:2001, Hofstetter:2004, Galpin:2005}  As a result of the Kondo effect, a sharp peak known as the Abrikosov-Suhl resonance\cite{Abrikosov-Suhl} (ASR) develops in the local density of states at the Fermi level. This ``Kondo peak'' enhances the conductance of the QD system when the electronic transport is limited by Coulomb blockade.	It has been recently demonstrated that the source  of the well-known 0.7 anomaly in quantum point contacts is of Kondo nature.\cite{Brun:2016}

	The Kondo effect in semiconductor artificial environments has been taken further to study more complex systems.  For example, the destructive interference between single and many-particle resonances, known as the Fano-Kondo effect, has been investigated in side-coupled quantum dots connected to different nanostructures.  These include connections to quantum wires,\cite{Kang:2001, Aligia:2002, Sato:2005} quantum rings,\cite{Bułka& Stefanski:2001, Hofstetter:2001, Kobayashi:2002} or to more quantum dots.\cite{Wu:2005, Tanamoto:2007, Sasaki:2009, Zitko:2010}   The rather complicated conductance features observed in these works emphasize the importance of the nontrivial interplay between coherence and many-body interactions.

	Other interesting Kondo-related physics manifest in nanostructures with multiple quantum dots.  For example, double\cite{Dias:2006, Logan:2009,Wong:2012} and triple\cite{Wang:2007, Mitchell:2009} quantum dots are known to exhibit a quantum phase transition (QPT) separating Strong-Coupling and Local-Moment phases.  In addition, the large number of geometrical configurations in triple quantum dot systems (TQD) permits the investigation of phenomena such as local frustration,\cite{Mitchell:2010, Seo:2013} ferromagnetic Kondo interactions,\cite{Baruselli:2013} and SU(4) behavior.\cite{Numata:2009} 
	
	Evenmore, the combination of interacting and effectively non-interacting dots in TQD allows the study of $e.g.$ the Kondo-Dicke effect,\cite{Trocha:2008} in which the ASR resonance is suppressed.\cite{Vernek:2010} 
There are other physical phenomena that may strongly alter   the structure of the ASR.  For example, a magnetic impurity inside a quantum corral\cite{Manoharan:2000} exhibits a splitting of its Kondo peak, depending upon the size of the enclosed section.\cite{Aligia:2001}  Aside from confinement, the splitting of the ASR can also be accomplished by applying an in-plane magnetic field to the Kondo system, for which the Zeeman energy is comparable to the Kondo temperature.\cite{Costi:2000} This can be understood in the context of the Anderson model, as a result of the spin-splitting of the impurity's energy level.\cite{Wright:2010} A similar behavior occurs in absence of external fields if the leads attached to a QD are ferromagnetic with parallel alignment between source and drain.\cite{Martinek:2003} An analog conclusion was reached in an experimental study of a single adatom coupled to a magnetic cluster, using the STM microscope.\cite{Kawahara:2010} Another different way to split the ASR in QDs is by applying a voltage bias. Here the splitting occurs due to the fact that the ASR is pinned to the Fermi level of each lead.\cite{Meir:1993} It has been noticed also that in a small QD exchange coupled to  a magnetic impurity,  the anisotropic exchange interaction between them gives rise to a structured ASR with more than two peaks.\cite{Tolea:2007} 

	A zero-field splitting of the Kondo peak may arise as well  in T-shaped double quantum dots (DQD).\cite{Dias:2006, Dias:2013} For this to happen, the QD connected to the leads must be essentially a resonant non-interacting level, while the ``side dot'' is set within the Kondo regime.  Although the splitting of the ASR can be interpreted here as a result of interference between single-particle and Kondo resonances,\cite{Aligia:2007} a more subtle explanation points towards a manifestation of the Friedel sum rule.\cite{Langreth:1966}  In general, this implies that the zero-temperature impurity's spectral function $A_{11}(\omega,T=0)$ evaluated at the Fermi level, must satisfy the condition: $A_{11}(0,0)=\sin^2(\pi\langle n_1\rangle/2+\phi)/[\pi \Delta(0)]$,\cite{Dias:2006, Dias:2013} where $\phi$ is a phase shift due to a structured hybridization function $\Delta(\omega)$ and $\langle n_1\rangle$ is the average of the impurity's number operator. In the aforementioned side-dot system, $1/[\pi \Delta(0)]$ provides an upper bound for the spectral function. In situations where $\Delta(\omega)$ causes $A_{11}(\omega)$ to lie above this boundary at either side of the Fermi level, the ASR develops a dip in order to fulfill the Friedel sum rule, and consequently displays a splitting.\cite{Dias-replay:2007} However, the splitting of the ASR due primarily to changes in the phase shift $\phi$, is still yet to be explored in detail.

	In this work, we use the Numerical Renormalization Group method to study the transport properties of a TQD that displays traces of the ASR zero-field splitting as described above, but attributable to a fine tuning of the phase shift $\phi$.  The device under investigation consists of two QDs, considered as effectively non-interacting, connected in parallel to metallic leads.  A third interacting dot (well within the Kondo regime) is only connected to the former dots.  We focus the paper to the study of two different configurations.  In the first, the on-site level of one of the two non-interacting dots is fixed at the Fermi energy. Calculations predict a splitting of the interacting-dot Kondo-peak by an appropriate tuning of the remaining non-interacting dot energy-level.  In a second configuration, one of the non-interacting dots is held out of resonance with the leads.  Under this scenario, the conductance exhibits a Fano-like shape, which is strongly modified by a quantum phase transition due to a pseudogap regime in the density of states of the effective one-impurity model. 

	The remainder of the paper is organized as follows: In Sec. \ref{sec:model} we describe the Hamiltonian of the TQD and present basic equations. Section \ref{kondosplit} is devoted to explore the conditions under which the interacting dot's spectral function displays a splitting at the Fermi level, as well as its identification through conductance measurements. In Sec. \ref{pseudogap} we show how the system can access a quantum phase transition, set by a pseudogap in the effective density of states. Concluding remarks appear in Sec. \ref{conclusions}.

\section{Hamiltonian and basic equations}
\label{sec:model}

\begin{figure}
\centerline{\includegraphics[width=0.9\columnwidth]{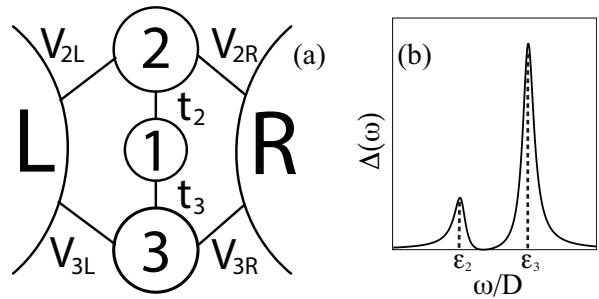}}
\caption{\label{Fig1}
(a) Schematic of the parallel triple quantum dot setup. Dots 2 and 3 are considered effectively non-interacting single particle levels, while dot 1 is deeply in the Kondo regime. (b) Illustrative representation of the effective density of states $\Delta(\omega)$ in the one impurity model.}
\end{figure}

We consider an equilibrium system represented schematically in Fig. \ref{Fig1} (a) and modeled by a generalized Anderson Hamiltonian

\begin{equation}
\label{H_full}
H = H_{\text{leads}}+H_{\text{dots}}+H_{\text{mix}}.
\end{equation}

\noindent The first term,

\begin{equation}
\label{H_leads}
H_{\text{leads}}=\sum_{\alpha,\bk,\sigma} \epsilon_{\alpha\bk}c_{\alpha\bk\sigma}^{\dag} c_{\alpha\bk\sigma}^{\pdag}
\end{equation}

\noindent represents the left ($L$) and right ($R$) metallic leads.  The operator $c_{\alpha\bk\sigma}$ annihilates an electron at lead $\alpha$ of wave vector $\bk$, with spin $z$ component $\sigma$ ($=\uparrow,\downarrow$), and energy $\epsilon_{\alpha\bk}$;

\begin{equation}
\label{H_dots}
H_{\text{dots}}=\sum_{i=1}^3 \varepsilon_i n_i+ U_1 n_{1\uparrow} n_{1\downarrow}+\sum_{i=2,\sigma}^3 t_i \bigl(
d_{1,\sigma}^{\dagger}d_{i,\sigma} + \text{H.c.} \bigr)
\end{equation}

\noindent describes the energetics of the dots in terms of their occupancies $n_{i\sigma}=d_{i\sigma}^{\dag} d_{i\sigma}^{\pdag}$ and $n_i=n_{i\uparrow}+n_{i\downarrow}$, where $d_{i\sigma}$ annihilates an electron of spin $z$-component $\sigma$ in the level $\varepsilon_i$ of the respective dot that lies closest to the common Fermi energy of the two leads (taken to be $\varepsilon_F =0$).  The parameter $t_i$ represents the dot $1$-dot $i$ ($i > 1$) coupling and $U_1$ characterizes the Coulomb strength at dot 1.  On the other hand, dots 2 and 3 are assumed to be single-particle levels, i.e. $U_2=U_3=0$.  One way to physically interpret this is to consider dots 2 and 3 as being close to a Coulomb blockade peak with single resonance-like behavior.  Finally,

\begin{equation}
\label{H_mix}
H_{\text{mix}}=\sum^3_{i=2,\alpha,\bk,\sigma} V_{\alpha,i} \bigl(d_{i\sigma}^{\dag}c_{\alpha\bk\sigma}+\text{H.c.} \bigr)
\end{equation}

\noindent accounts for electron tunneling between dots and leads. For simplicity, we take real dot-lead couplings $V_{iL}=V_{iR}\equiv V_i/\sqrt{2}$, such that the dots interact only with one effective band formed by an even-parity combination of $L$ and $R$ states. We assume a constant density of states $\rho=1/(2D)$ with half bandwidth $D$, so that the dot-lead tunneling is measured via the hybridization widths $\Gamma_i=\pi\rho V_i^2$.  

The standard equations of motion $\omega \langle\langle d_{m}^{\pdag}; d_{n}^{\dag}\rangle\rangle_{\omega} -\langle[d_m,d_n^{\dagger}]\rangle  =  \langle\langle[d_m,H];d_n^{\dagger}\rangle\rangle_{\omega} = -\langle\langle d_m;[d_n^{\dagger},H]\rangle\rangle_{\omega}$ for the retarded Green's function  $\langle\langle d_m;d_n^{\dagger}\rangle\rangle =-i\int_0^{\infty}dt e^{i\omega t}\langle\{d_{m}^{\pdag}(t), \, d_{n}^{\dag}(0)\}\rangle \equiv G_{mn}(\omega)$, allow to estimate (see Appendix) the dot-1 Green function as $G_{11}(\omega) = [1+U\Theta(\omega)]G^{(0)}_{11}(\omega)$. Here, $G^{(0)}_{11}(\omega)$ corresponds to the non-interacting function in presence of the other two dots, $\Theta(\omega)$ is an extra contribution to $G^{(0)}_{11}$ due to the electron-electron interactions and $\omega$ is the energy measured from the chemical potential. In the wide-band limit ($D\gg |\omega|$), the non-interacting dot-1 Green function can be written as $[G^{(0)}_{11}(\omega)]^{-1}=\omega_1-\Lambda(\omega)+i \Delta(\omega)$, where we have defined $\omega_i=\omega-\varepsilon_i$ and

\begin{equation}
\Delta(\omega)=\lambda(\omega)(\omega_2t_3\sqrt{\Gamma_3}+\omega_3t_2\sqrt{\Gamma_2})^2, 
\end{equation}

\begin{align}
\Lambda(\omega)&=\lambda(\omega)[\omega_2\omega_3(\omega_3t_2^2+\omega_2t_3^2)\nonumber \\
&+(\omega_2\Gamma_3+\omega_3\Gamma_2)(t_2\sqrt{\Gamma_3}-t_3\sqrt{\Gamma_2})^2]
\end{align}

\noindent with $\lambda(\omega)=1/[(\omega_2\omega_3)^2+(\omega_2\Gamma_3+\omega_3\Gamma_2)^2]$. The coupling of dot 1 to the leads can be extracted from the Green function $G_{11}(\omega)$.  For our model, this information is fully contained in the functions $\Lambda(\omega)$, which only shifts the single particle energy $\varepsilon_1$, and more importantly on $\Delta(\omega)$.  This provides a mapping of the Hamiltonian \eqref{H_full} onto the single-impurity Anderson model, in which the impurity is coupled to an electronic band via an effective hybridization function $\Delta(\omega)$.  Therefore, it is possible to numerically treat the problem as a single-impurity model.  

Figure 1 (b) provides an illustrative version of $\Delta(\omega)$.  This effective density of states has two Lorentzian shapes, centered near $\varepsilon_2$ and $\varepsilon_3$, respectively.  The width and height of each Lorentzian is proportional to the parameters $t_i$ and $\Gamma_i$.  However, it must be pointed out that the function $\Delta(\omega)$ goes beyond the simple sum of two Lorentzian functions.  For any $\varepsilon_2\ne\varepsilon_3$, this effective density of states vanishes at $\omega=(\varepsilon_3\sqrt{\Gamma_2}t_2+\varepsilon_2\sqrt{\Gamma_3}t_3)/(\sqrt{\Gamma_2}t_2+\sqrt{\Gamma_3}t_3)$.  If this value coincides with the Fermi energy, $\Delta(\omega)$ displays features of a pseudogapped host.\cite{pseudogap, Ingersent-pseudo}   In this regime, the Strong-Coupling fixed point is inaccessible under particle-hole symmetry, but the system exhibits quantum criticallity as a function of the interacting-dot energy level.\cite{Dias:2006}  This is an important result of our work, and it is explored further in this paper.

At low bias, electron transmission though the TQD can be described by a Landauer-like formula,\cite{Meir:92} giving a linear conductance

\begin{equation}\label{g}
g(T)=g_0\int d\omega \frac{-\partial
f(\omega,T)}{\partial\omega}
 [-\text{Im}{\cal T}(\omega,T)]
\end{equation}

\noindent where $g_0=2e^2/h$ is the maximum of conductance,  $f(\omega,T) = [\exp(\omega/T)+1]^{-1}$ is the Fermi-Dirac distribution at temperature $T$ and ${\cal T}(\omega,T) = \sum_{m,n}\sqrt{\Gamma_{m}\Gamma_n}G_{mn}(\omega,T)$ is the transmission function.  For our particular system, all the Green's functions $G_{mn}$ can be expressed in terms of $G_{11}(\omega,T)$ via the Equation of Motion Technique (see Appendix), resulting in

\begin{align}\label{TransFunc}
-\text{Im}{\cal T}(\omega)&=[2\omega_2^2\omega_3^2\lambda(\omega)-1]\pi\Delta(\omega)A_{11}(\omega,T) \nonumber \\ 
&+ 2\omega_2\omega_3(\omega_3\Gamma_2+\omega_2\Gamma_3)\lambda(\omega)\Delta(\omega)G^{\prime}_{11}(\omega,T) \nonumber  \\
&+1-\omega_2^2\omega_3^2\lambda(\omega).
\end{align}

\noindent Here, $A_{11}(\omega,T)=-\pi^{-1} \, \text{Im} \,G_{11}(\omega,T)$ is the dot-1 spectral density.  The real part of the Green function $G^{\prime}_{11}(\omega,T) = \text{Re}G_{11}(\omega,T)$ can be computed by making a Hilbert transform of $A_{11}(\omega)$.  At zero temperature, eq. \eqref{g} reduces to $g/g_0=-\text{Im}{\cal T}(0)$, allowing to obtain analytical expressions for the electrical conductance.

	The Fermi energy properties of the spectral density are governed by a Fermi-liquid relation known as the Friedel sum rule (FSR).  In systems featuring a non-trivially structured density of states, the spectral function must satisfy\cite{Dias:2013}

\begin{equation}\label{FSR}
\pi\Delta(0)A_{11}(0,0)=\sin^2(\pi\ \langle n_i\rangle/2+\phi),
\end{equation}

\noindent where the phase $\phi$ is given by

\begin{equation}\label{phasePhi}
\phi=\text{Im}\int_{-\infty}^0\frac{\ \partial\Sigma_{11}^0(\omega,T=0)}{\ \partial \omega}G_{11}(\omega,T=0)d\omega
\end{equation}

\noindent being $\Sigma_{11}^0(\omega,T=0)=\Lambda(\omega)-i\Delta(\omega)$ the non-interacting impurity self-energy.  As we will show, the phase $\phi$ plays a crucial role in the transport properties of the TQD system.

	The study of the effective single-impurity model has been carried out using the Numerical Renormalization Group method.\cite{Bulla:08} To this aim, we have employed a discretization parameter $\Lambda_{NRG}=2.5$, retaining at least 1000 states after each iteration.  In the remaining of the paper, dot 1 is held at the Coulomb blockade regime, with $U_1=0.4D$ and $\varepsilon_1=-U/2$.  Without loss of generalization, we have fixed the values of the hybridization strength to $\Gamma_2=\Gamma_3=0.02D$ and $t_2=0.02D$ as well.  We focus on the transport properties of the system, as a function of the dots' energy levels $\varepsilon_i$'s, which should be experimentally tunable via plunger gate voltages. 

\section{Splitting of the Abrikosov Suhl resonance.}
\label{kondosplit}

	Before starting the numerical analysis, it results instructive to consider the properties of a fully symmetric TQD device, namely $t_2=t_3=t$, $V_{\alpha, 2}=V_{\alpha,3}=V_{\alpha}$ and $\varepsilon_2=\varepsilon_3=\varepsilon$.  Taking even ($e$) and odd ($o$) combinations of dot levels 2 and 3, $i.e.$ $d_{e,\sigma}=(d_{2,\sigma}+d_{3,\sigma})/\sqrt{2}$ and $d_{o,\sigma}= (d_{2,\sigma}-d_{3,\sigma})/\sqrt{2}$, the Hamiltonian \eqref{H_full} becomes

\begin{align}\label{effectiveDQD}
H_{\text{dots}}+H_{\text{mix}}&=\varepsilon_1 n_1+ U_1 n_{1\downarrow}n_{1\uparrow}+\varepsilon (n_{e}+n_{o})\nonumber\\
& + \sqrt{2}\sum_{\sigma} (t d_{1,\sigma}^{\dagger}d_{e,\sigma} + \text{H.c})\nonumber\\
& + \sqrt{2}\sum_{\alpha,\bk,\sigma} ( V_{\alpha} d_{e\sigma}^{\dag}c_{\alpha\bk\sigma}+\text{H.c.}),
\end{align}

\noindent and $H_{\text{leads}}$ remains unchanged.  Therefore, only the $e$-orbital couples directly to both, the leads and dot 1. In other words, the Hamiltonian reduces to a T-shaped DQD device with dot 1 as the``side dot".  For small values of $t$ such that $T_K\ll\Gamma$, the interacting-quantum-dot spectral function exhibits the usual Kondo peak structure, $i.e.$  $\langle n_1\rangle=1$, $\phi=0$ and $\pi\Delta(0)A_{11}(0,0)=1$.  We expect the same behavior for $\varepsilon_2\ne\varepsilon_3$, if $t$ is small enough.

	On the other hand, when $\varepsilon_2=0$, the zero-temperature conductance is simply given by $g/g_0 = 1-\pi\Delta(0)A_{11}(0,0)$, and $\Delta(\omega=0,\varepsilon_2=0)=t_2^2/\Gamma_2$ is independent of $\varepsilon_3$ and $t_3$.  For the symmetric scenario with small $t$ depicted above, no transport as a function of $\varepsilon_3$ should occur when $T$ vanishes.  Hence, in presence of Kondo correlations, the ASR allows an extra conduction path through dot 1 that gives rise to destructive interference with dot 2, leading to zero transport through the TQD. 

\begin{figure}[b]
\centerline{\includegraphics[width=0.7\columnwidth]{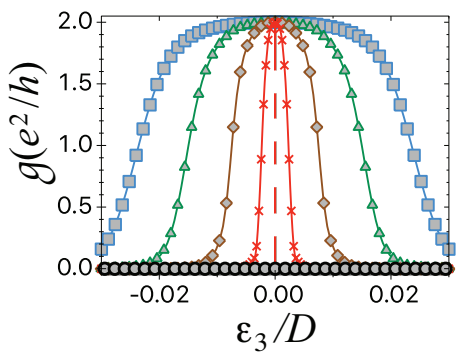}}
\caption{\label{Fig2}
(Color online) Zero-temperature conductance vs $\varepsilon_3$ for $\varepsilon_2=0$ and several values of $t_3/D$: 0.02 (black $\bullet$), 0.03 (red $\times$), 0.04 (brown $\blacklozenge$), 0.05 (green $\blacktriangle$), 0.06 (blue $\blacksquare$). For $t_3=t_2$ (black $\bullet$), $g=0$ regardless $\varepsilon_3$.  However, when $t_3\neq t_2$, a proper adjustment of $\varepsilon_3$ leads to a finite conductance, due to a splitting in the dot-1 ASR.}
\end{figure}

\begin{figure}
\centerline{\includegraphics[width=0.9\columnwidth]{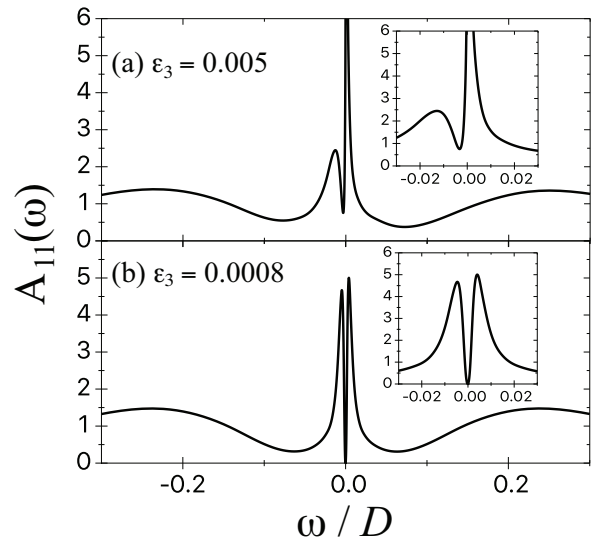}}
\caption{\label{Fig3}
Zero-temperature spectral function for (a) $\varepsilon_3=0.005$ and (b) $\varepsilon_3=0.0008$.  Here $t_3=0.04D$ and the rest of the parameters as in Fig. \ref{Fig2}. The insets at each panel are amplifications of the respective spectral function, near the Fermi energy. As shown in (b), the spectral function nearly vanishes at $\omega=0$ even in presence of the Kondo effect, in order to fulfill a generalized Friedel sum rule.}
\end{figure}

	The preceding discussion is now considered in Fig. \ref{Fig2}.  Here we plotted $g/g_0$ vs $\varepsilon_3$ for different values of $t_3$.  We first focus on the symmetric case $t_3=t_2$  (black $\bullet$) which as discussed above, exhibits $g/g_0=0$ independently of $\varepsilon_3$.  However, this is no longer true for the asymmetric case $t_3\neq t_2$.  As shown by Fig. \ref{Fig2}, increasing $t_3$ above $t_2$ opens up two windows of finite conductance at each side of $\varepsilon_3=0$.  The range of $\varepsilon_3$ values for which the conductance is non-zero, can be increased either by increasing $t_3$ above $t_2$ (as shown by Fig. \ref{Fig2}) or by decreasing $t_3$ below $t_2$ down to zero (not shown).   The conductance curves exhibit reflection symmetry at $\varepsilon_3=0$, since at this point the system is particle-hole symmetric.  For this reason, $g(\varepsilon_3=0)=0$, regardless the value of $t_3$.  

	The finite conductance behavior for the asymmetric case $t_3\neq t_2$ is one of the main results of this paper.   As previously discussed, the contributions to the non-zero conductance at $T=0$ must come from changes in the ASR such that $\sin^2(\pi\ \langle n_1\rangle/2+\phi)\neq1$.  To verify this, in Fig. \ref{Fig3} we have plotted $A_{11}(\omega,T=0)$ for $t_3/D=0.04$ and different values of $\varepsilon_3$ such that $g/g_0\neq0$.  In both cases, the spectral function away from the Fermi level exhibits the usual Hubbard bands near $\omega=\varepsilon_1$ and $\omega=\varepsilon_1+U$.  Also both panels show a large peak in the vicinity of $\omega=0$, that can be identified as the ASR.  To better appreciate the behavior of the spectral density at the Fermi level, we have included as insets, an amplification of $A_{11}(\omega)$ for each case.  In both figures, a splitting of the ASR is clearly appreciated.  More importantly, for the case depicted in panel (b) ($\varepsilon_3=0.0008$), the spectral function at the Fermi level is almost zero.  We emphasize that in the case at hand, $\Delta(\omega=0)=t_2^2/\Gamma_2$ is a constant finite value.  Therefore, the vanishing of the spectral function at the Fermi level, does not come from $1/[\pi \Delta(0)]$ providing an upper bound for $A_{11}(0)$, but  from $\pi\ \langle n_1\rangle/2+\phi$ being close to a multiple integer of $\pi$.  

\begin{figure}
\centerline{\includegraphics[width=0.98\columnwidth]{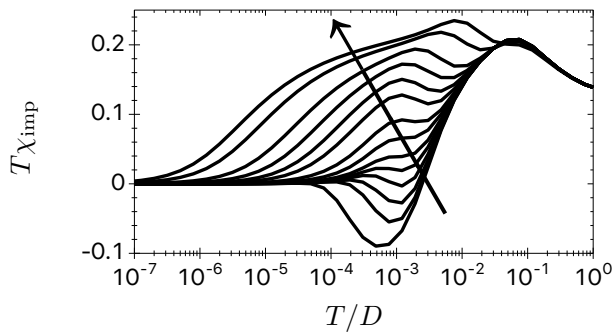}}
\caption{\label{Fig4}
(Color Online) Spin susceptibility as a function of temperature for various $\varepsilon_3$ (increasing in the direction of the arrow) in the range $[0.001,0.07]$.  Here $t_3=0.04D$ and the rest of the parameters as in Fig.\ref{Fig2}. In all curves, $T\chi_{\text{imp}}$ vanishes with the temperature, congruent with Kondo physics.}
\end{figure}

	To corroborate that the Kondo correlations persist whenever the ASR is splitted, in Fig. \ref{Fig4} we plotted the impurity's contribution to the susceptibility as a function of the temperature $T\chi_{\text{imp}}(T)$. This quantity is calculated in the standard fashion\cite{Bulla:08, Krishna-murthy:80} as $T\chi_{\text{imp}}(T)=(\langle S_z^2 \rangle - \langle S_z \rangle^2)-(\langle S_z^2 \rangle - \langle S_z \rangle^2)_0$, where $S_z$ is the $z$ component of the total system spin. The symbol $\langle...\rangle$ denotes thermal average and the subscript $0$ refers to the situation when no impurities are present.  In this figure, $t_3=0.04D$ and the rest of the parameters as in Fig. \ref{Fig2}.  We have plotted $T\chi_{\text{imp}}$ for several values of $\varepsilon_3$ such that $0.001\le \varepsilon_3/D\le0.07$.  It is evident that all curves vanish with the temperature. Therefore the Kondo ground state is reached, independently of $\varepsilon_3$. On the other hand, the curves exhibit a diamagnetic behavior\cite{Dias:2006, Mastrogiuseppe:2014, Zhuravlev:2011, Hofstetter:1999} ($\chi_{imp}<0$) for $\varepsilon_3/D\lesssim0.004$.  It has been recently pointed out that such result, for a density of states finite at the Fermi level, is a spurious outcome from the traditional NRG method\cite{Feng:2015} and that the susceptibility curves should follow the universal shape. This asseveration supports our claim that the ground state exhibited by our TQD is indeed a Kondo singlet. For $\varepsilon_3\gtrsim0.004$, the manner in which $T\chi_{imp}\to 0$ as $T \to 0$, is identical to that in the Kondo regime of the conventional Anderson model.

\begin{figure}[b]
\centerline{\includegraphics[width=0.95\columnwidth]{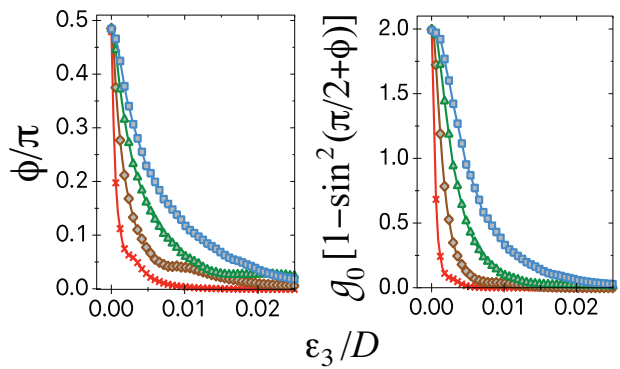}}
\caption{\label{Fig5}
(Color Online) (a) Phase $\phi$ and (b) $g_0[1-\sin^2(\pi\ \langle n_1\rangle/2+\phi)]$ vs $\varepsilon_3$, for $\varepsilon_2=0$ and several values of $t_3/D$: 0.03 (red $\times$), 0.04 (brown $\blacklozenge$), 0.05 (green $\blacktriangle$), 0.06 (blue $\blacksquare$). This result shows that the splitting of the ASR can be attributable to a fine tuning of the phase $\phi$.}
\end{figure}

	In order to better support the description of the splitted ASR, in Fig. \ref{Fig5} (a) we have numerically calculated the phase $\phi$ as a function of $\varepsilon_3$, from equation \eqref{phasePhi}.  It is important to point out that the accurate computation of the phase is a very challenging task.  On one hand, one needs to calculate the real part of the Green's function by taking a Hilbert transform of the spectral function, which in turn contains errors due to NRG discretization and {\it ad hoc} broadening procedures.  On the other hand, for the system described in this paper, the derivative of the non-interacting impurity self energy $\partial\Sigma_{11}(\omega)/\partial \omega$ has strong variations near $\omega=\varepsilon_3$.   Given the limited information that the NRG provides about the spectral function and since the integral in \eqref{phasePhi} must be evaluated within the range $[-\infty,0]$ in energy, we only show the parameter space corresponding to $\varepsilon_3>0$. 

	Despite all the inconveniences inherent to the numerical evaluation of $\phi$, Fig. \ref{Fig5} (a) provides a clear confirmation on how the generalized FSR eq. \eqref{FSR} explains the splitting of the ASR, and consequently the transport properties of the TQD.  As a comparison to Fig. \ref{Fig2} we also show in Fig. \ref{Fig5} (b) the conductance computed directly from the expression $1-\sin^2(\langle n_1 \rangle \pi /2+\phi)$. For all the curves shown, $\langle n_1\rangle$ is practically  $1$. Although there is no quantitative agreement between Fig. \ref{Fig2} and Fig. \ref{Fig5} (b), the last one certainly displays the expected trend.  Also, note in Fig. \ref{Fig5} (a),  that the phase is $\phi \lesssim \pi/2$ close to $\varepsilon_3=0$, which entails a vanishing intensity of the ASR at the Fermi level, and a high conductance.  In addition, at a given $\varepsilon_3$ value such that the phase is finite, both $\phi$ and the conductance increase with $t_3$.  Lastly, the two properties also decrease monotonically down to zero, as $\varepsilon_3$ is tuned away from the Fermi level.

\begin{figure}[b]
\centerline{\includegraphics[width=0.98\columnwidth]{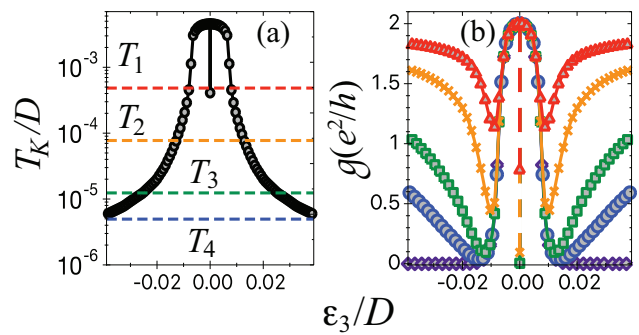}}
\caption{\label{Fig6}
(Color Online) (a) Kondo temperature $T_K$  and (b) finite-$T$ conductance $g$ vs $\varepsilon_3$.  The horizontal lines in (a) correspond to the temperatures in (b) at which the conductance was calculated: $T=0$ (purple $\blacklozenge$), $T_1$ (blue  $\bullet$), $T_2$ (green $\blacksquare$), $T_3$ (orange $\times$) and $T_4$ (red $\blacktriangle$).  Here $t_3=0.04D$ and the rest of the parameters as in Fig.\ref{Fig2}. As shown by the plots, the phase $\phi$ can be experimentally determined through electrical conductance measurements.}
\end{figure}

	Interestingly, the conductance signatures of the ASR splitting persist at $T>0$.  However, to better understand the temperature dependence of $g/g_0$, it is instructive to consider first the evolution of the Kondo temperature with $\varepsilon_3$.  This is depicted in Fig. \ref{Fig6} (a).  Here we consider $t_3=0.04D$ and the rest of the parameters as in Fig. \ref{Fig2}.  The Kondo temperature $T_K$ was determined via the conventional criterion $T_K \chi_{imp}(T_K) = 0.0701$.\cite{note} As shown by the graphic, $T_K$ decreases asymptotically to $T/D=1.16\times10^{-6}$ by increasing $|\varepsilon_3|$.  This value corresponds to the Kondo temperature in absence of dot 3.  Decreasing $|\varepsilon_3|$ results in an increment of states in the vicinity of the Fermi level, favoring the Kondo correlations and therefore increasing $T_K$.

	Finite-$T$ conductance calculations are shown in Fig. \ref{Fig6} (b).  The color code corresponds to the four temperatures $T_1<T_2<T_3<T_4$, shown as horizontal lines in Fig. \ref{Fig6} (a). In panel (b) we also include the $T=0$ curve.  We recall that for large values of $|\varepsilon_3|$, the spectral function has a regular ASR, with no splitting.  Increasing the temperature above $T_K$ gradually destroys the Kondo resonance.  This in turns limits the access through dot 1, as well as the respective destructive interference given by the different transport paths.  As a result, the conductance through the TQD system is increased. On the other hand, note that for small values of $\varepsilon_3$, $g/g_0$ remains almost unchanged with respect to the $T=0$ case, due to the high values of $T_K$.  We stress here that a non-zero conductance behavior in the vicinity of $\varepsilon_3=0$ constitutes the finger print of the ASR splitting.  Therefore, the corresponding magnitude of the phase $\phi$ may be inferred from conductance measurements at experimentally accessible temperatures.

\section{Quantum phase transition and Fano-Kondo effect.}
\label{pseudogap}

	In this section we consider a more general case $\varepsilon_2\ne0$.  When the dot-2 energy-level is away from resonance with the leads, the electronic transport between source and drain takes place either directly through dot 3 or indirectly through dot 1.  This gives rise to Fano-Kondo interference.  On the other hand, in this configuration the effective one-impurity model hybridization function exhibits a pseudogap behavior whenever $\varepsilon_3\ne\varepsilon_2$ and $\varepsilon_3=-\varepsilon_2 t_3 \sqrt{\Gamma_2 \Gamma_3}/\Gamma_2 t_2\equiv\varepsilon^*$.  In this scenario, $\Delta(\omega)$ vanishes as $\omega^2$ near the Fermi energy.  The Kondo effect in presence of pseudogapped density of states has been widely studied.\cite{Dias:2006,pseudogap,Ingersent-pseudo}  It is well established that it exhibits a QPT, between Kondo and non-Kondo phases.  Therefore, the TQD proposed in this paper allows a systematic study of the Fano-Kondo effect in the presence of a QPT, scenario to our knowledge, not considered before.  This is the main focus of this section.

\begin{figure}
\centerline{\includegraphics[width=0.9\columnwidth]{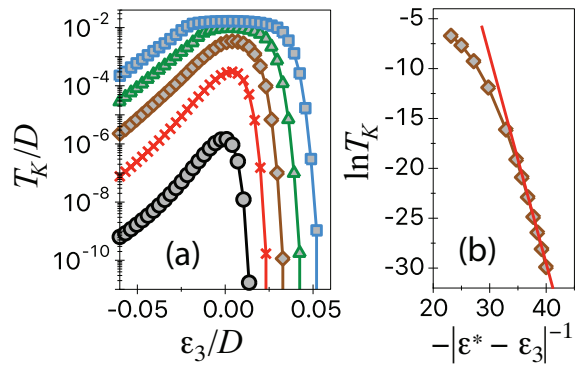}}
\caption{\label{Fig7}
(Color online). (a) $T_K$ vs $\varepsilon_3$,  for several values of $t_3/D$: 0.02 (black $\bullet$), 0.03 (red $\times$), 0.04 (brown $\blacklozenge$), 0.05 (green $\blacktriangle$), 0.06 (blue $\blacksquare$). (b) $\ln T_K$ vs $-|\varepsilon^*-\varepsilon_3|^{-1}$, for $t_3/D=0.04$. In both figures, $\varepsilon_2=-0.03D$.  The behavior of $T_K$ against $\varepsilon_3$ is characteristic of a Kosterlitz-Thouless quantum phase transition.}
\end{figure}

	We start by analyzing the dependence of $T_K$ on $\varepsilon_3$.  This is shown in Fig. \ref{Fig7} (a), for $\varepsilon_2=-0.03D$ and several values of $t_3$.  We point out that for finite $\varepsilon_2$, $T\chi_{\text{imp}}$ does not exhibit a diamagnetic behavior (not shown).  As in section \ref{kondosplit}, the behavior of $T_K$ can be explained in terms of the hybridization function $\Delta(\omega)$. For a given $\varepsilon_3$, $T_K$ increases with $t_3$, as it happens with the magnitude of $\Delta(0)$.  At $\varepsilon_3=0$, the hybridization function has a Lorentzian centered at the Fermi level, and the Kondo temperature acquires its maximum value.  If $\varepsilon_3$ is further increased towards $\varepsilon^*$,  the system approaches to the pseudogap regime, and $T_K$ rapidly vanishes.  This behavior is characteristic of a quantum phase transition (QPT), separating Strong-Coupling (Kondo) and Local-Moment (non-Kondo) phases.   This is a second important result of this paper.  To better understand the nature of the QPT, in Fig. \ref{Fig7} (b) we have plotted $\ln T_K$ vs $-|\varepsilon^*-\varepsilon_3|^{-1}$, for the particular case $t_3/D=0.04$.  As shown by the graphic, there is a linear dependence between these two quantities, when $\varepsilon_3\to\varepsilon^*$.  This is confirmed by the red line shown for eye-guide.  Therefore, the QPT is of the Kosterlitz-Thouless type,\cite{KT-transition} in which $T_K$ vanishes exponentially, rather than as a power law. Note that the system exhibits an unquenched behavior only at the transition point, $\varepsilon_3=\varepsilon^*$.

	To better illustrate the implications of the QPT on the transport properties of the TQD, we consider again a symmetric system, $i.e.$ $t_2=t_3=t$. In this case, the contribution from the second term in eq. \eqref{TransFunc} to $-\text{Im}{\cal T}(0)$ can be neglected, since $G^{\prime}_{11}(0)$ is very small.  Moreover, due to the fact that the phase $\phi$ is expected to vanish, one can also assume $\pi\Delta(0)A_{11}(0)=1$.  Collectively, these considerations reduce the zero-temperature conductance to the following Fano-like expression: 

\begin{equation}\label{SymZeroTG}
g/g_0=\frac{\ (\varepsilon_2 \varepsilon_3)^2}{\ (\varepsilon_2 \varepsilon_3)^2+(\varepsilon_2\Gamma_3+\varepsilon_3\Gamma_2)^2}.
\end{equation}
	
\noindent We remind here that eq. \eqref{SymZeroTG} is valid for $\varepsilon_3\ne \varepsilon^*$. In the opposite case and as long as $\varepsilon_1=-U/2$, the dot-1 spectral function vanishes at the Fermi level, $i.e.$ Kondo correlations are prevented as a consequence of the pseudogapped host.  This in turn is detrimental to the Fano-Kondo interaction as we shall further discuss in the following.  Then, for $\varepsilon_3= \varepsilon^*$, $g/g_0=1-(\varepsilon_2\varepsilon_3)^2\lambda(0)$, which thus gives zero transport for $\Gamma_2=\Gamma_3$.

	The analysis of the conductance in a more general asymmetric case ($t_2\ne t_3$) can not be carried out analytically and a numerical evaluation of eq. \eqref{TransFunc} is required.  In this scenario, the phase $\phi$ is now expected to be finite. Evenmore, there is no reason to presume a negligible contribution to the conductance from the term involving $G^{\prime}_{11}(0)$.  Another important point of consideration is the behavior of the Kondo temperature near the pseudogap regime, which as we showed, vanishes exponentially with $|\varepsilon^*-\varepsilon_3|^{-1}$.  Strictly speaking, the calculation of the $T=0$ properties of any system through the NRG methodology requires an infinite number of numerical iterations.  Being this unfeasible, it is a common practice to stop the iterative procedure once the stable fixed point has been reached. Within the Strong-Coupling fixed point, this occurs when the temperature $T_N\sim\Lambda_{NRG}^{-(N-1)/2}$, associated to the $N$-th iteration, is below $T_K$. However, the Kondo temperature decreases exponentially near the pseudogap regime ($\varepsilon_3\to \varepsilon^*$) in the case under study.  For this reason, we have chosen to show conductance calculations at $T=T^{\prime}=2.13\times10^{-12}D$, equivalent to $N=60$ iterations with discretization parameter $\Lambda_{NRG}=2.5$.  Typically, this temperature corresponds to nano-Kelvins.

\begin{figure}[b]
\centerline{\includegraphics[width=\columnwidth]{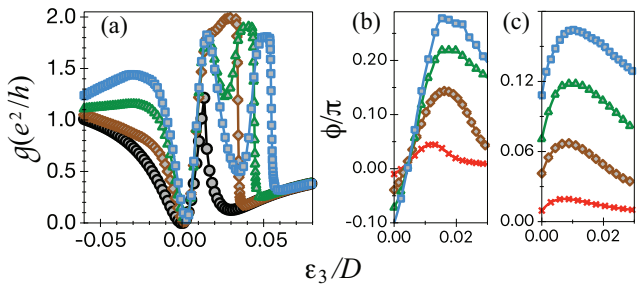}}
\caption{\label{Fig8}
(Color online). (a) $g$ vs $\varepsilon_3$, for $\varepsilon_2/D=-0.03$.  Phase $\phi$ vs $\varepsilon_3$ for (b) $\varepsilon_2/D=-0.03$ and (c) $\varepsilon_2/D=0.03$.  .  In all three panels, we have several values of $t_3/D$: 0.02 (black $\bullet$), 0.04 (brown $\blacklozenge$), 0.05 (green $\blacktriangle$), 0.06 (blue $\blacksquare$). The drastic drop in $g$ at $\varepsilon_3>0$ is a result of the quantum phase transition.}
\end{figure}

	Figure \ref{Fig8} (a) shows $g$ vs $\varepsilon_3$, for $\varepsilon_2/D=-0.03$ and various values of $t_3$.  For a given $\varepsilon_3$ such that $T_K>T^{\prime}$ (see Fig. \ref{Fig7} (a)), the behavior of $g$ is essentially that of the $T=0$ case, which follows a Fano-like shape.  Since dot-2 is out of resonance with the leads, the TQD electronic's transport occurs primarily through dots 1 and 3. In fact, when $\varepsilon_3$ is tuned near to the Fermi energy, the electrical conductance vanishes independently of $t_3$.  This obeys to destructive Fano-Kondo interference between transport through dots 1 and 3.  When $\varepsilon_3$ is increased above the Fermi level, the transport takes place only through dot 1 (assisted by the``Kondo cloud''), raising $g$. We note however, that for $e.g.$ $t_3=0.06D$ (blue $\blacksquare$), the conductance exhibits a dip at $\varepsilon_3 \approx 0.035$.  It is easy to show that around this value, $-\text{Im}{\cal T}(0)\approx \pi\Delta(0)A_{11}(0)$.  Hence, the minimum observed in $g$ results from the spectral function fulfilling the FSR eq. \eqref{FSR}.

	On the other hand, if $\varepsilon_3$ is such that $T_K<T^{\prime}$, thermal effects associated to the QPT deviate the conductance curves from the Fano shape.  In all four curves, we appreciate a drastic drop of the conductance, when $\varepsilon_3$ is further increased near their respective transition values $\varepsilon^*$.  This is a manifestation of the QPT.  We recall that the calculation has been performed considering a very low, but finite temperature.  Therefore, the pseudogap regime does not only shows itself right at $\varepsilon^*$, but there are traces of this behavior in the vicinity of the transition value.  In this regime, no Kondo effect occurs.  Evenmore, the on-site energy of both non-interacting dots are away from resonance with the leads.  All these together redound in a low conductance behavior.

	Finally, in figs. \ref{Fig8} (b) and (c) we show the dependence of the phase $\phi$ on $\varepsilon_3$ for $\varepsilon_2/D=-0.03$ and $0.03$ (respectively) calculated directly form eq. \eqref{phasePhi}. The results are shown for different values of $t_3$.  As in section \ref{kondosplit}, we only display results for the range $\varepsilon_3>0$.  Nevertheless, given the symmetry of the system it is possible to infer the $\varepsilon_3<0$ behavior of $\phi$, for (lets say) $\varepsilon_2=-0.03$,  by taking the results from the case $\varepsilon_2=0.03$ and making $\phi\to-\phi$ and $\varepsilon_3\to-\varepsilon_3$.  In both panels [figs. \ref{Fig8} (b) and (c)], the phase $\phi$ increases with $t_3$ and vanishes for large values of $|\varepsilon_3|$, similarly to the $\varepsilon_2=0$ case.  However, now $\phi$ presents a maximum at a finite $\varepsilon_3$. Given the complicated expression for the $T=0$ conductance when $\varepsilon_2 \ne 0$, it is not possible to relate the behavior of $\phi$ to $g$ for an arbitrary $\varepsilon_3$.  Nevertheless, these results show that the dot-1 spectral function still exhibits a splitting of the ASR, whenever $\phi$ is finite.

\section{Conclusion}
\label{conclusions}

	In summary, the transport properties of a triple quantum dot system have been investigated.  The device under consideration consists of two effectively non-interacting dots connected in parallel to metallic leads, as well as to a central interacting quantum dot.  The described arrangement permits a comprehensive study of the Kondo-resonance zero-field splitting, resulting from the restrictions imposed by the Friedel sum rule generalized to systems with structured density of states.  This splitting can be tuned by an appropriate change of the non-interacting quantum dots on-site energy levels.   When both non-interacting dots are nearly resonant with the metallic leads, the interacting dot spectral function at the Fermi level approaches to zero in presence of Kondo correlations.  On the other hand, when one non-interacting dot energy-level is away from the Fermi energy, the conductance exhibits traces of Fano-Kondo interference.  Also in this configuration, a quantum phase transition (of the Kosterlitz-Thouless type) inherent to a pseudogap behavior, strongly affects the Fano-Kondo effect.  All these behaviors can be inferred from experimental measurements.	
	
\begin{acknowledgments}
We thank Kevin Ingersent and Sergio Ulloa for helpful discussions. We also acknowledge supercomputing resources received from UNAM and DGTIC to develop this work, through the program ``Miztli'' 2016, project SC15-1-S-70.  AW thanks funding from Conacyt, through the National Posdoctoral Program. FM acknowledges support from the project Dynamics of and Complex Systems: 612707-FP7-PEOPLE-2013-IRSES.
\end{acknowledgments}

\appendix
\section{Equation of motion}

In this appendix, we use the Equation of Motion technique to calculate the dot-1 Green function, as described in section \ref{sec:model}.  In general, the retarded Green function

\begin{equation}
\langle\langle d_i;d_j^{\dagger}\rangle\rangle =-i\int_0^{\infty}dt e^{i\omega t}\langle\{d_{i}^{\pdag}(t), \, d_{j}^{\dag}(0)\}\rangle \equiv G_{ij}(\omega)
\end{equation}

\noindent obeys equations of motion of the form

\begin{align}\label{EOM}
\omega G_{ij}(\omega)-\langle[d_i,d_j^{\dagger}]\rangle &= \langle\langle[d_i,H];d_j^{\dagger}\rangle\rangle \\
&= -\langle\langle d_i;[d_j^{\dagger},H]\rangle\rangle.
\end{align}

For the Hamiltonian \eqref{H_full}

\begin{multline}
[d_{i,\sigma},H]=(\varepsilon_i+U_i n_{i,-\sigma})d_{i,\sigma}+\delta_{i1}\sum_lt_ld_{l,\sigma}\\ +\sum_l\delta_{il}t_ld_{1,\sigma}+\sum_{\alpha,\bk}V_{\alpha,i}c_{\alpha\bk\sigma},
\end{multline}

\noindent being $\delta_{ij}$ the Kronecker delta.  Using

\begin{equation}
\langle\langle c_{\alpha\bk\sigma};d_{i,\sigma}^{\dagger}\rangle\rangle=\frac{\ 1}{\ \omega-\epsilon_{\alpha,\bk}}\sum_lV_{\alpha,l}G_{li}(\omega),
\end{equation}

\noindent which can also be deduced from the equation of motion \eqref{EOM}, it is easy to arrive to the following general expression for the retarded Green function:

\begin{multline}\label{GreenOne}
G_{ij}(\omega)=G^{(0)}_i\bigr\{\delta_{ij}+U_i\langle\langle n_{i,-\sigma}d_{i,\sigma};d_{j,\sigma}^{\dagger}\rangle\rangle + \delta_{i1}\sum_lt_lG_{lj}(\omega) \\
+\sum_l \delta_{il} t_l G_{1j}(\omega)-i\sum_{l \ne i}\Gamma_{il}G_{lj}(\omega)\bigr\},
\end{multline}

\noindent where $G^{(0)}_i=(\omega_i+i\Gamma_i)^{-1}$ is the non-interacting Green function of dot $i$ in absence of the other dots, $\omega_i=\omega-\varepsilon_i$ and $\Gamma_{ij}=\sqrt{\Gamma_i \Gamma_j}$.  Note that we have obtained \eqref{GreenOne} within the wide band approximation.  Following a similar procedure, it is possible to obtain

\begin{multline}\label{GreenTwo}
G_{ij}(\omega)=G^{(0)}_j\bigr\{\delta_{ij}+U_j\langle\langle d_{i,\sigma}; n_{j,-\sigma}d_{j,\sigma}^{\dagger}\rangle\rangle + \delta_{1j}\sum_lt_lG_{il}(\omega) \\
+\sum_l \delta_{lj} t_l G_{i1}(\omega)-i\sum_{l \ne j}\Gamma_{jl}G_{il}(\omega)\bigr\}.
\end{multline}

With the aid of \eqref{GreenOne} and \eqref{GreenTwo} we explicitly get

\begin{align}
G_{11}(\omega)/G_1^{(0)}&=1+U_1\Theta(\omega)+t_2G_{21}+t_3G_{31}\\
G_{21}(\omega)/G_2^{(0)}&=t_2G_{11}-i\Gamma_{23}G_{31}\\
G_{31}(\omega)/G_3^{(0)}&=t_3G_{11}-i\Gamma_{32}G_{21}
\end{align}

\noindent where $\Theta(\omega)=\langle\langle n_{1,-\sigma}d_{1,\sigma};d_{1,\sigma}^{\dagger}\rangle\rangle$.  The above system of equations can be solved for $G_{11}$ yielding the interacting dot-1 Green function as described in section \ref{sec:model}.  In the same fashion, it is possible to explicitly obtain the transmission function ${\cal T}(\omega)$, in terms of the interacting Green function $G_{11}$ as written in \eqref{TransFunc}.

\end{document}